\documentclass[aps,prb,twocolumn,groupedaddress,showpacs,amsmath,amssymb]{revtex4-1}
\usepackage{dcolumn}   
\usepackage{amsmath}
\usepackage{graphicx}    
\usepackage{subfigure}
\usepackage{color}
\usepackage{comment}

\begin{document}

\setlength{\parskip}{1.5mm}


\title{Fermi surfaces and Phase Stability of Ba(Fe$_\text{1-x}$\emph{M}$_\text{x}$)$_2$­As$_2$ (\emph{M}=Co, Ni, Cu, Zn)}
\author{S. N. Khan,$^{1,2}$ Aftab Alam,$^{3}$ and Duane D. Johnson$^{1,2,4}$}
\email{snkhan@illinois.edu,ddj@ameslab.gov,aftab@phy.iitb.ac.in}
\affiliation{$^{1}$Department of Physics, University of Illinois, Urbana-Champaign, Illinois 61801, USA;} 
\affiliation{$^{2}$The Ames Laboratory, US Department of Energy, Ames, Iowa 50011-3020, USA;}
\affiliation{$^{3}$Department of Physics, Indian Institute of Technology, Bombay, Powai, Mumbai 400 076, India}
\affiliation{$^{4}$Department of Materials Science \& Engineering, Iowa State University, Ames, Iowa 50011-2300.}


\begin{abstract}
BaFe$_2$As$_2$ with transition-metal doping exhibits a variety of rich phenomenon from coupling of structure, magnetism, and superconductivity. Using density functional theory, we systematically compare the Fermi surfaces (FS), formation energies ($\Delta E_f$), and density of states (DOS) of electron-doped Ba(Fe$_{1-x}$M$_{x}$)$_2$As$_2$ with M=\{Co, Ni, Cu, Zn\} in tetragonal (I$4/mmm$) and orthorhombic (F$mmm$) structures in nonmagnetic (NM), antiferromagnetic (AFM), and paramagnetic (PM, disordered local moment) states. 
We explain changes to phase stability ($\Delta E_f$) and Fermi surfaces (and nesting) due to chemical and magnetic disorder, and compare to observed/assessed properties and contrast alloy theory with that expected from rigid-band model. 
With alloying, the DOS changes from common-band (Co,Ni) to split-band (Cu,Zn), which dictates $\Delta E_f$ and can overwhelm FS-nesting instabilities, as for Cu,Zn cases. 
\end{abstract} 
\date{\today}
\maketitle


\section{Introduction}
The Fe-based superconductors (Fe-SCs) are providing new avenues to explore high-T$_c$ superconductivity (SC) involving magnetism.\cite{RevModPhys.83.1589,LowTempPhy.38.888,SupercondSciTech.23.073001,NatPhys.6.645,AnnualRevCMP.2.121,AnnualRevCMP.1.27} BaFe$_2$As$_2$ (BFA) has been of particular interest due to its ease of synthesis. The parent compound exhibits AFM order at a Ne{\'e}l temperature of \mbox{T$_\text{N}$ = 140 K} and this order is suppressed in favor of SC on chemical doping.\cite{PhysRevLett.101.117004,PhysRevB.79.014506,PhysRevB.82.024519,PhysRevB.78.214515} Electron doping can be achieved by substituting a transition metal for Fe to give a metal-substituted solid solution M-BFA,\cite{PhysRevB.78.214515,PhysRevB.82.024519} with an increasing electron-per-Fe ($e$/Fe) count. Notably, different behaviors for chemical and magnetic ordering instabilities are found for Co,Ni versus Cu,Zn solid-solutions, for example, with (in)commensurate AFM order depending on the dopant, with possible coexistence of SC and AFM order. Interestingly, Cr-based binary, metallic alloys\cite{PhysRevB.34.3446,0305-4608-8-10-007} show the same coexistence behavior but with T$_c$ an order of magnitude smaller than Fe-SCs. Although numerous experimental studies on electron-doped BFA have been carried out over the last few years, a systematic theoretical investigation is still lacking. Here, we address Ba(Fe$_\text{1-x}$M$_\text{x}$)$_2$As$_2$  via a proper alloy theory to provide a direct comparison of trends and explain their origin.
 
To provide a single theoretical description within density functional theory (DFT), we use KKR multiple-scattering theory combined with the coherent-potential approximation (CPA)\cite{PhysRevLett.56.2088} to handle chemical and magnetic disorder.
For M-BFA in \mbox{high-T} I$4/mmm$ and \mbox{low-T} F$mmm$ structures\cite{PhysRevB.78.020503} with increasing $e$/Fe in the NM, PM, and AFM states, the KKR-CPA is used to examine relative phase stability ($\Delta E_f$), Fermi surface topologies and nesting (electron-hole) features through the Bloch spectral functions,\cite{0305-4608-15-8-007} and changes of the density of states (DOS) due to alloying and disorder, as well as to contrast these results to expectations from a rigid-band model.

\section{Background}
Generally, DFT results on BFA match the striped AFM ordering\cite{PhysRevB.79.184523} and measured electronic structure quite well. The FS exhibits two or three hole cylinders at the zone center ($\Gamma$) and two electron cylinders at the zone corner (X), as observed in DFT\cite{PhysRevB.78.094511} and angle-resolved photoemission\cite{PhysRevB.79.155118,PhysRevLett.104.137001,JPSJ.78.123706} (ARPES). The prominent ($\pi$,$\pi$) FS nesting between these cylinders helps stabilize the AFM state,\cite{LowTempPhy.38.888} and spin fluctuations in this mode may drive Cooper pairing.\cite{PhysRevB.82.054515,PhysRevB.78.195114} Hence, the need to study FS nesting and disorder broadening effects.

The M-BFA phase diagrams show suppressed AFM ordering in favor of a neighboring SC state. The SC domes have M-fraction x=0.03-0.12, 0.02-0.08, and $\sim$0.04 with T$^{max}_c$ of  23, 20, and 2~K for Co-, Ni-, and Cu-BFA, respectively.\cite{PhysRevB.79.014506,PhysRevB.82.024519} Zn-doped samples do not superconduct. Notably, T$^{max}_c$ occurs near the extrapolated AFM quench concentration.\cite{PhysRevB.79.014506,PhysRevB.82.024519} For Co- and Ni-BFA, the magnetic order becomes an incommensurate spin-density wave before entering the SC state, which emphasizes itinerancy and import of FS nesting.\cite{PhysRevLett.109.167003} Cu-BFA remains commensurate.\cite{PhysRevLett.109.167003} And, no changes in the magnetism or FS are found in Zn-BFA.\cite{PhysRevB.87.201110} In addition, there are steric effects due to changing $a$ and $c$ lattice constants in I$4/mmm$ structure. $a$ is almost unchanged for Co-BFA and increases for Ni-, Cu-, and Zn-BFA. For Co-, Ni-, Cu-BFA $c$ shrinks\cite{PhysRevB.79.024512,PhysRevB.79.153102,PhysRevLett.109.097002} and for Zn-BFA it increases.\cite{Xiao2013} 

Lastly, there is debate on whether M-BFA follows a rigid-band picture; and, if not, whether an effective Fermi level shift is still applicable. In a rigid-band model, the electronic structure is fixed to that of BaFe$_2$As$_2$, and the Fermi energy is raised by the amount of additional $e$/Fe for each dopant, as determined by their atomic number $Z_i$: Co (1$e$), Ni (2$e$), Cu (3$e$), and Zn (4$e$); the atomic species of the dopant becomes irrelevant (even though $Z_i$ increases and changes the scattering properties relative to Fe), and all dopants should generate the same electronic effects for a given $e$/Fe. As such, a proper alloy theory can make clear assessments. While ARPES shows similar trends with nominal $e$/Fe for Co- and Ni-BFA, there are deviations from rigid-band for Cu- and Zn-BFA.\cite{PhysRevLett.110.107007,PhysRevB.87.201110} By Luttinger's theorem\cite{PhysRev.119.1153}, an effective $e$/Fe can be defined from changes in the experimentally measured FS. The phase diagrams of Co-, Ni-, and Cu-BFA have been found to approximately coincide in this manner.\cite{PhysRevLett.110.107007} Zn-BFA shows no measurable changes in FS and no superconductivity.\cite{PhysRevB.87.201110} Comparing supercell calculations for Co- and Zn-BFA show that Co-BFA obeys rigid-band while Zn-BFA does not.\cite{PhysRevLett.108.207003} The rigid-band model is applicable as long as site-potential differences between Fe and dopant are much less than bandwidths. As we show, these differences are visible in a dopant's site-projected DOS, where significant overlap between Fe $d$-states and those of Co or Ni exists, less so for Cu, and almost none for Zn.\cite{PhysRevLett.105.157004} We find that the FS evolves similar to that expected from rigid-band for Co-, Ni-, and Cu-BFA but not for Zn-BFA; yet, deviations from rigid-band behavior are readily apparent in $\Delta E_f$ for PM Cu-BFA. 



\section{Computational Details}

DFT calculations at 0~K were performed using an all-electron, KKR-CPA Green's function method.\cite{Physica.13.392,PhysRev.94.1111,PhysRevLett.56.2088} To improve the usual site-centered basis set, empty spheres (E1, E2, and E3) were inserted at interstitial voids in the structure (Table \ref{spheres}). A local density approximation to DFT is used and the coherent potential approximation (CPA) is used to address chemical and magnetic disorder.\cite{PhysRevB.82.024435} 
For PM states, uncorrelated, randomly-oriented local moments (site magnetizations $m_i\ne0$) are described by a disordered local moment (DLM) state,\cite{0305-4608-15-6-018} where such site magnetic disorder can produce large energy broadening of the electronic states, which is reduced with magnetic short-range order included (beyond the CPA\cite{PhysRevB.72.113105}), but changes FS nesting contributions to the magnetic susceptibility.\cite{PhysRevLett.50.374}  The DLM state  is often a more appropriate representation of the PM state than the NM state ($m_i=0$) typically assumed in theory for comparison to experiment,\cite{PhysRevB.85.220503,Yin2011} such as for magnetic transition temperatures in magnetic metals.\cite{PhysRevB.66.014416,PhysRevB.82.024435} 
 
All results were obtained with a 8$\times$8$\times$8 Monkhorst-Pack \mbox{k-point} mesh for Brillouin zone (BZ) integrals,\cite{PhysRevB.13.5188} and using complex energy (E) contour integration with 25 E-points on a \mbox{Gauss-Legendre} semi-circular contour.\cite{PhysRevB.30.5508} 
Fermi energies were determined from an analytic, integrated DOS (Lloyd's) formula\cite{0953-8984-16-36-011} to yield an accurate electron count.
The valence configurations were taken as \mbox{Ba 5$p^6$$6s^2$}, \mbox{Fe 4$s^2$$3d^6$}, \mbox{Co 4$s^2$$3d^7$}, \mbox{Ni 4$s^2$$3d^8$}, \mbox{Cu 4$s^2$$3d^9$}, \mbox{Zn 4$s^2$$3d^{10}$}, and \mbox{As 4$s^2$$4p^3$}. 
To match the $e$-per-volume of the BaFe$_2$As$_2$ samples probed in experiment, lattice constants (in $pm$) were fixed to experiment:\cite{PhysRevB.78.020503} (I$4/mmm$) \mbox{$a$ = $b$ = 396.25} and \mbox{$c$ = 1301.68}, and (F$mmm$) \mbox{$a$ = 561.46}, \mbox{$b$ = 557.42}, and \mbox{$c$ = 1294.53}. 
As the alloy concentrations are sufficiently low, we fixed the lattice to minimize DFT (relative) error and isolate electronic and steric effects.

\begin{table}
{\caption{Atomic coordinates and sphere sizes for atoms and empty spheres (E1-3). I$4/mmm$ (F$mmm$) has body-centered tetragonal (face-centered orthorhomic) unit vectors.}
\begin{tabular}[b]{cccc}
\hline
Site & Coordinates & Wyckoff & Radius ($pm$)\\ \hline
I$4/mmm$ & & & \\
Ba & $(0.0000a, 0.0000a, 0.0000c)$ & 2a & 225.1 \\ 
Fe(M) & $(0.5000a, 0.0000a, 0.0000c)$ & 4d & 136.5 \\ 
As & $(0.0000a, 0.0000a, 0.3545c)$ & 4e & 136.5 \\ 
E1 &  $(0.5000a, 0.5000a, 0.0000c)$ & 2b & 76.3 \\ 
E2 & $(0.0000a, 0.0000a, 0.2072c)$ & 4e & 78.9 \\
E3 & $(0.2007a, 0.2007a, 0.1715c)$ & 16m & 55.8 \\
\\
F$mmm$ & & & \\
Ba & $(0.0000a, 0.0000b, 0.0000c)$ & 4a & 224.4 \\ 
Fe(M) & $(0.2500a, 0.2500b, 0.2500c)$ & 8f & 136.1\\ 
As & $(0.0000a, 0.0000b, 0.3545c)$ & 8i & 136.1 \\ 
E1 &  $(0.5000a, 0.0000b, 0.0000c)$ & 4b & 76.1 \\ 
E2 & $(0.0000a, 0.0000b, 0.2072c)$ & 8i & 78.6 \\
E3 & $(0.2007a, 0.0000b, 0.1715c)$ & 16n & 55.7\\ 
\hline
\label{spheres}
\end{tabular}
}
\end{table}

Fermi surfaces were determined at E$_F$ via the Bloch spectral function $A({\bf k},E) = -\frac{1}{\pi}\,\text{Im}\,G({\bf k},E)$, where $G$ is the single-particle Green's function. $A({\bf k},E)$ is the $E$- and {\bf k}-space resolved DOS and dispersion. In the limit of an ordered compound it reduces to Dirac \mbox{$\delta$-functions} that define the band structure $E({\bf k})$. In the presence of magnetic or chemical disorder there is {\bf k}-dependent spectral broadening and shifting due to impurity scattering handled via the CPA. The spectral \mbox{full-width} at \mbox{half-maximum} with respect to energy is inversely proportional to the lifetime of electronic states,\cite{0305-4608-10-12-007,PhysRevB.29.4217} which also dictate transport and SC properties. Spectral broadening also can support coexistence of AFM and SC, as found, for example, in binary Cr alloys, such as Cr-Ru.\cite{PhysRevB.34.3446}



\begin{figure}[t]
\centering
\includegraphics[width=9cm]{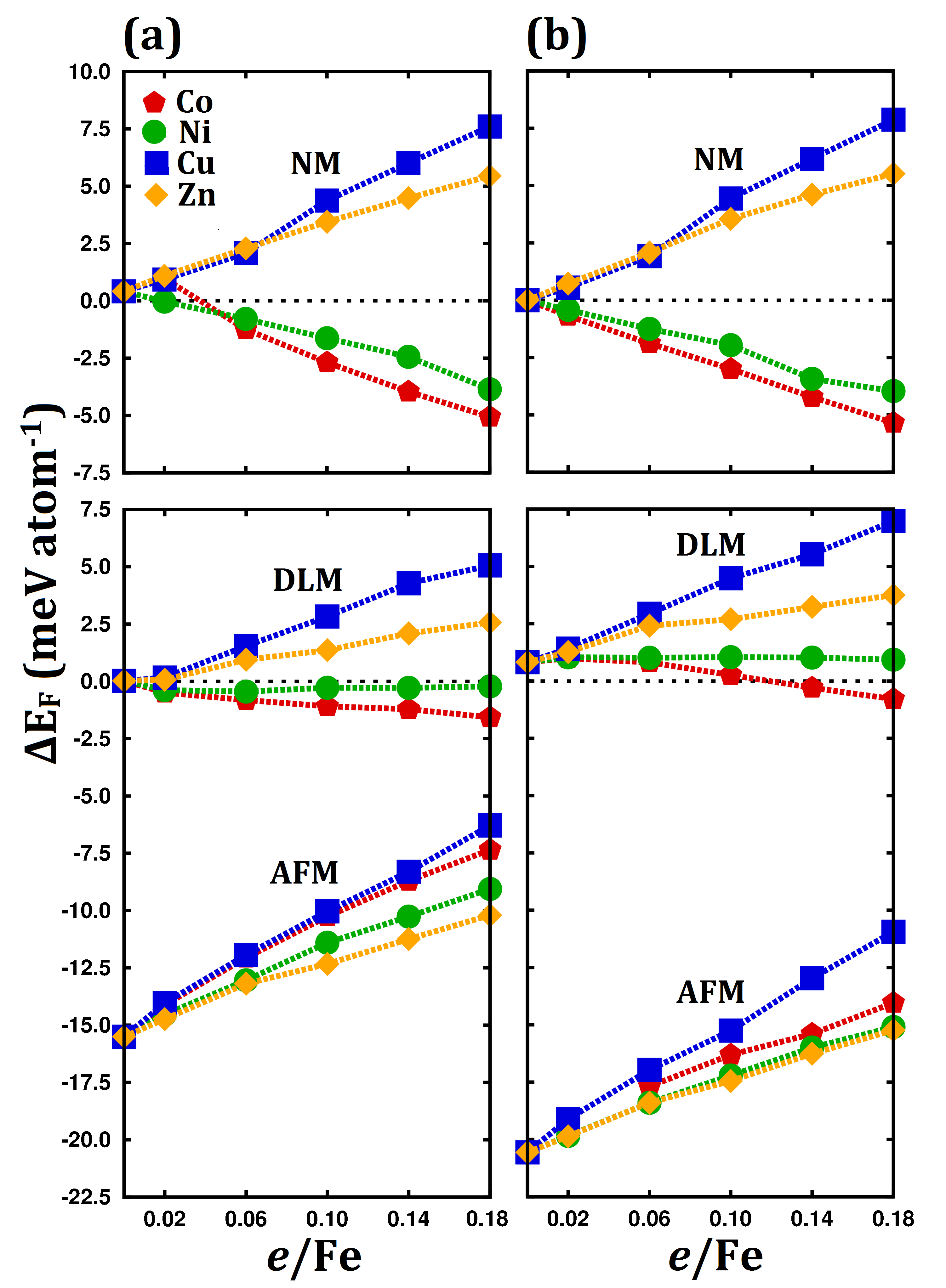}
\caption{(Color online) Formation energy of NM, DLM, and AFM Ba(Fe$_\text{1-x}$\emph{M}$_\text{x}$)$_2$­As$_2$ relative to mixed phase NM endpoints BaFe$_2$As$_2$ and Ba$M_2$As$_2$  in (a) I$4/mmm$ and (b) F$mmm$ structural phases. Nominal $e$/Fe counting is used.}
\label{fig-energetics}
\end{figure}

\section{Results}

\subsection{Phase Stability with Alloying}
For an alloy, the formation energies is defined relative to the concentration-weighted sum of the energy of the (Ba,Fe,M,As) constituents in their respective equilibrium structures.
However, in a fixed host, $\Delta{E}_f$ trends for Ba(Fe$_{1-x}$M$_{x}$)$_2$As$_2$ versus $x$ (or $e$/Fe) are more simply revealed referencing BaFe$_2$As$_2$ and BaM$_2$As$_2$, i.e.,
\begin{eqnarray}\label{Eform}
\Delta{E}_f = && E^{\text{Ba(Fe}_{1-x}\text{M}_{x})_{2}\text{As}_{2}} \nonumber  \\ 
              &-& [({1-x})E^{\text{BaFe}_{2}\text{As}_{2}}+ {x} E^{\text{BaM}_{2}\text{As}_{2}} ].
\end{eqnarray}
Figure~\ref{fig-energetics} shows $\Delta E_f$ of the NM, DLM, and AFM states versus nominal $e$/Fe for each M, plotted relative to the mixed phase with NM BaFe$_2$As$_2$ and MFe$_2$As$_2$. 

For no doping, the NM and DLM energies are nearly degenerate. At finite temperatures the DLM state will have a lower free-energy due to spin-disorder entropy. The AFM state is 16 (I$4/mmm$) or 21 $me$V/atom (F$mmm$) below the NM state.  In experiments on BFA, the magnetic and structural phase transition occur simultaneously in BFA at 140 K (or 12 $me$V).\cite{PhysRevB.79.014506} Previous DFT studies find 37 $me$V/atom (I$4/mmm$) using full-potential augmented plane waves (FLAPW)\cite{PhysRevB.78.094511} or 70 $me$V/atom (F$mmm$) using plane-wave pseudopotentials (PWP),\cite{PhysRevB.79.184523} which is the available data.

For magnetism, we find Fe site moments (I$4/mmm$) are 1.4 $\mu_\text{B}$ (AFM) and 1.0 $\mu_\text{B}$ (DLM). For F$mmm$, there is only a slight drop to 1.3 $\mu_\text{B}$ (AFM) and 0.95 $\mu_\text{B}$ (DLM). We do not find symmetry breaking to lead to a significant change of site moments. As a contrast, we note local, Heisenberg models, often fit to spin-wave spectra, find very different J$_{1a}$ and J$_{1b}$ nearest neighbor exchange parameters with broken symmetry.\cite{PhysRevB.84.054544,Zhao2009} We see, however, that magnetic disorder leads to a large reduction in site moments, due to transverse components of the magnetization. The time-averaged ordered moment is 0.9 $\mu_\text{B}$ from neutron diffraction,\cite{PhysRevLett.101.257003} which have some transverse components. Using core-electron spectroscopy to probe short-time scales (10$^{-15}$ sec), the measured moment is 2.1 $\mu_\text{B}$ in the closely related SrFe$_2$As$_2$.\cite{PhysRevB.85.220503} This difference has been attributed to modest electron correlations\cite{Yin2011} and magnetic excitations.\cite{Mazin09} We see here that the disordered component of the site moment is substantial. Overall, the KKR results agree reasonably with previously computed AFM site moments of 1.8 $\mu_\text{B}$ (I$4/mmm$) from FLAPW; the moments from PWP are 2.6 $\mu_\text{B}$ (F$mmm$). 

In this low-doping regime ($e$/Fe $\leq$ 0.18), the $\Delta{E}_f$ vary linearly with $x$ or $e$/Fe for all magnetic states and structures. Furthermore, the resulting trends are robust whether considering the I$4/mmm$ or F$mmm$ structures. In the NM state, there is a clear splitting in the behavior of Co- and Ni-BFA versus Cu- and Zn-BFA. Both Co- and Ni-BFA show the same, favorable formation energies for given $e$/Fe. Cu- and Zn-BFA also agree for given $e$/Fe but are unfavorable to mixing at zero temperature. 
Chemical mixing entropy does reduce formation enthalpies relative to endpoints BaFe$_2$As$_2$ and Ba$M_2$As$_2$, increasing the favorability of the higher $e$/Fe compounds. In an ideal mixing model\cite{note.entropy} this will reduce the free energy by 21, 13, 10, and 8 $me$V/atom for Co-, Ni, Cu-, and Zn-BFA at $e$/Fe = 0.18 and 1000\,$^\circ$C, a typical annealing temperature.\cite{PhysRevB.79.014506,PhysRevB.82.024519} This effect is not accounted for in the 0~K results in Fig.~\ref{fig-energetics} so as to separate electronic ($e$/Fe) effects from entropic (dopant \mbox{$x$}) effects. 

\begin{figure*}[]
\includegraphics[width=18cm]{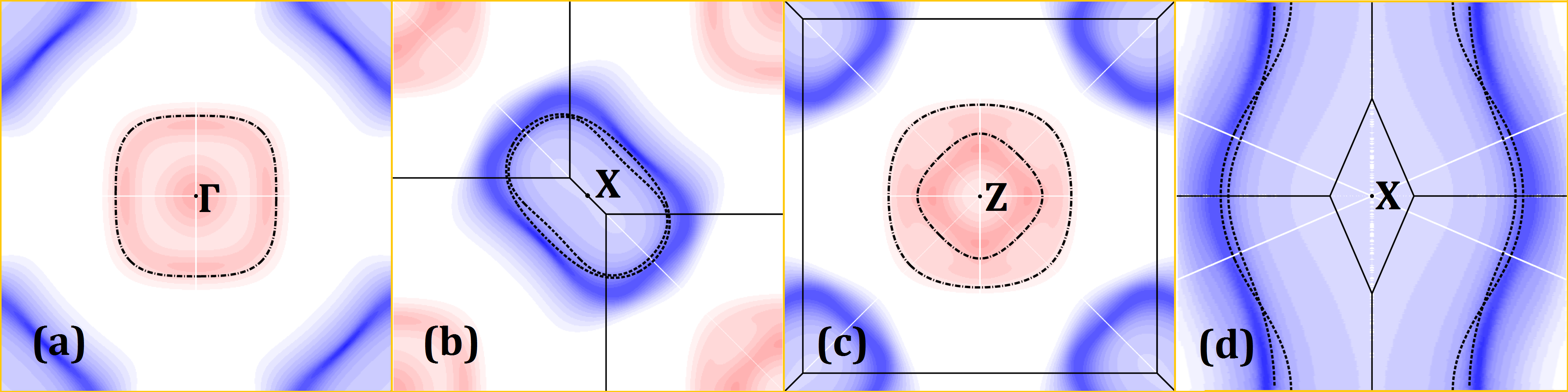}
\caption
{(Color online) Bloch spectral function $A({\bf k},E)$  of BaFe$_2$As$_2$ in both NM and DLM states. Hole cylinders are depicted as dash-dot lines (NM) or false-color scaled red (DLM). Electron cylinders are dashed (NM) or blue (DLM). Solid lines indicate the BZ boundary. Cross-sections are normal to (a) [001] about $\Gamma$, (b) [001] about X, (c) [001] about Z, and (d) [$\bar{1}$10] about X. False coloring is mapped (in 10$^3$ states Ryd$^\text{-1}$cell$^\text{-1}$ r.u.$^\text{-3}$) as \{0 $\leftrightarrow$ RGB 0xFFFFFF (white), 100 $\leftrightarrow$ RGB 0x770000 (light red), and 1000 $\leftrightarrow$ RGB 0xFF0000 (red)\} for hole pockets. And similarly for electron pockets. This choice was made to make as many features as possible visible across plots. }
\label{fig-purefermi}
\end{figure*}

In the DLM state, a similar splitting persists, but energies are less pronounced. There is also less agreement in the energies of Cu- and Zn-BFA at a given $e$/Fe. We find no magnetic moments at the dopant atom and only marginally reducing moments on the Fe sites with increasing $e$/Fe. Neutron diffraction shows a rapid drop in Fe moment with doping.\cite{PhysRevLett.109.167003} This may be a result of the sensitivity of the moment to an increasing $a$ lattice constant.\cite{PhysRevLett.101.126401} Experiments that demonstrate the incommensurability of the spin-density wave\cite{PhysRevLett.109.167003} on Co and Ni doping are done in the PM state. Cu-BFA does not become incommensurate. Our results show Cu mixing is, at best, weakly favorable. Thus, the lack of incommensurate splitting in Cu might arise not as a result of a FS effects, but rather due Cu clustering. In the AFM state, dopants decrease favorability relative to the PM state. This is in qualitative agreement with the known phase diagrams, where dopants suppress the AFM state and eventually lead to SC. The dopant species splitting here is even less pronounced and all compounds follow nearly the same trend with $e$/Fe. This suggests an important difference in doping effects on the PM and AFM state. Note that prior DFT calculations for the doped compounds have been performed on the NM state.\cite{PhysRevLett.105.157004,PhysRevLett.108.207003}


\subsection{Fermi Surfaces of PM States}
Figure~\ref{fig-purefermi} shows the FS of NM and DLM BaFe$_2$As$_2$. NM surfaces are shown for electrons (dashed lines) or holes (dash-dot lines)  -- there is no FS broadening with no chemical disorder -- and these surfaces agree with previous results. The Brillouin zone (solid lines) and labels correspond to the body-centered tetragonal lattice and can be found in the literature.\cite{Setyawan2010299} 
DLM  surfaces for electrons (blue) or holes (red) show significant broadening due to local orientational disorder -- in contrast to chemical disorder, which we see, below, is less significant. The approximate {\bf k}-space broadening is $0.14~r.u.$ (reciprocal units defined as $2\pi/a$ units in ${\bf k}$-space). 

Note that DLM Bloch spectral peaks do not coincide exactly with the NM surface. The DLM hole (electron) pockets are reduced (enlarged) in size relative to the NM. This corresponds to an effective $e$-doping, as reflected in the DOS with a positive shift of $E_F$. The interior pocket near the Z point is pinched off near the $\Gamma$ point. This can vary with choice of exchange-corrleation and lattice parameters. A strong pinching is also visible in prior DFT calculations\cite{PhysRevB.78.094511} and ARPES\cite{JPSJ.78.123706,PhysRevLett.110.107007}. The outer cylinder is fairly uniform and gives rise to strong nesting with electron cylinders. The electron cylinders obey a 4$_1$ screw symmetry along the k$_\text{z}$-axis while the hole cylinders obey $90\,^{\circ}$ rotational symmetry. The DLM broadening and $E_F$ shift changes the strength of nesting between hole and electron cylinders. The large broadening can explain the reduced resolution of ARPES data, especially when compared to measurements made on CuO SCs.

To make a connection to nesting, we note that it is, in principle, possible to calculate the chemical, magnetic and magneto-chemical susceptibilities within the KKR-CPA using a thermodynamic linear-response theory,\cite{PhysRevLett.65.1259} similar to phonon linear-response that uses infinitesimal displacements. For such susceptibilities in the high-symmetry (disordered) state, the correct functional form is $\chi^{-1}({\bf q};T) \sim [1 - \beta (1-m_i^2) S^{(2)}({\bf q},T) ] $, where $\beta=(k_{B}T)^{-1}$ and $S^{(2)}({\bf q},T)$ is an exact second-variation of the electronic grand-potential with respect to fluctuations, e.g., site magnetizations.
For an Ising-like system, $S^{(2)}({\bf q},T)$ plays the role of a thermodynamically averaged pairwise $J({\bf q})$. Such calculations have been done for solid-solutions\cite{PhysRevLett.65.1259,PhysRevLett.66.766,PhysRevLett.74.138,PhysRevLett.74.3225} and elemental FM\cite{PhysRevLett.69.371} and AFM,\cite{PhysRevLett.82.3340} but not yet for multi-sublattice cases. 

\begin{figure*}[]
\includegraphics[width=18cm,height=5cm]{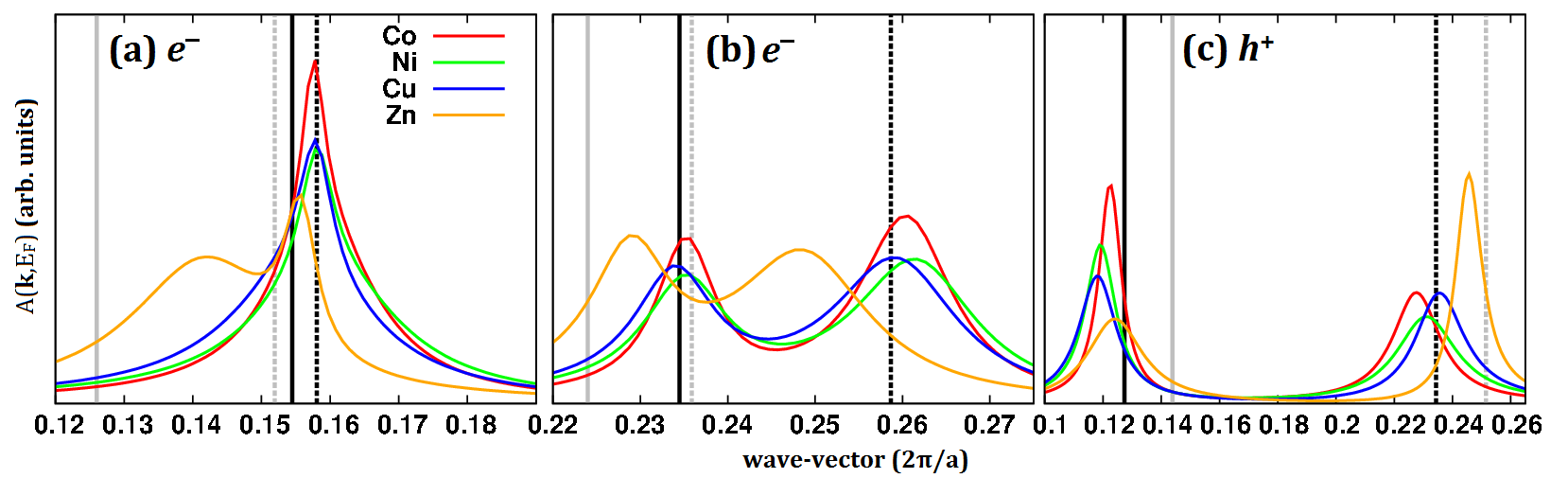}
\caption
{(Color online) For NM Co-, Ni-, Cu-, and Zn-BFA at $e$/Fe = 0.10, the Bloch-spectral function $A({\bf k},E)$ along (a) X=[$\frac{1}{2}\frac{1}{2}0$] to $\Gamma$=[000] (electron), (b) X to Z=[010] (electron), and (c) Z to X (hole). These correspond to traversing the principal axes about the electron and hole cylinders. Black vertical lines correspond to rigid-band expectation at the same $e$/Fe: first (second) line is the inner (outer) cylinder. First (second) gray line is for the inner (outer) cylinder of parent BFA. Clearly, Zn is behaving differently than Co, Ni, Cu, or the rigid-band. In (c), hole states deviate notably from rigid-band behavior.}
\label{fig-spanv}
\end{figure*}

Nonetheless, $S^{(2)}({\bf q},T)$, with  matrix elements $M(\epsilon)$ and Fermi factor $f(\epsilon)$, is a generalized susceptibility:
\begin{eqnarray}
S^{(2)}({\bf q};T) &\sim& \int d\epsilon~M(\epsilon) \int d\epsilon' \left[\frac{f(\epsilon;T)-f(\epsilon';T)}{\epsilon-\epsilon'}\right] \nonumber \\ 
 &\times& \frac{1}{\Omega_{BZ}}\int~d{\bf k}~A({\bf k};\epsilon)A({\bf k}+{\bf q};\epsilon')  \label{S2} \\
 &\rightarrow&  \int~d{\bf k}~A({\bf k};E_F)A({\bf k}+{\bf q};E_F) . \label{S2nest}
\end{eqnarray}
In principle, all states in the valence contribute to \eqref{S2}. If only hole and electron states near $E_F$ dominate, the bracketed factor $[...]$ yields \eqref{S2nest}, which is a convolution of the Fermi surface states and the origin for ``nesting''.\cite{PhysRevLett.50.374,PhysRevLett.74.138}  Due to alloying, even in a metallic system, hybridized states well below $E_F$ can drive ordering (NiPt\cite{PhysRevLett.66.766}) or only features at $E_F$ (CuPt\cite{PhysRevLett.74.3225}). For Cr, the NM state yields nesting with a incommensurate wavevector, as observed,\cite{RevModPhys.66.25} while, for Cr-Ru the chemical disorder broadens the FS enough that the SDW now is commensurate, as observed, and coexists with SC.

This discussion was to motivate that the DLM state (with similar FS topology to the NM state) typically creates similar nesting due to larger volumes of the Brillouin zone contributing to the susceptibility integral, even though the peak overlap is reduced.


\subsection{Fermi-Surface Nesting in NM State}

We analyze the NM Fermi surface (electrons and holes) typically used for SDW stability analysis for a given $e$/Fe. A cross-section of $A({\bf k},E_\text{F})$ for transition-metal alloys at fixed $e$/Fe = 0.10 are shown in Fig.~\ref{fig-spanv} (in $r.u.$), which traverses from the center of electron (hole) cylinders along principal axes  k$_1$ = [110] and k$_2$ = [$\bar{1}$10]. Only a range near the spectral peaks is shown in each case. The k-space broadening is $\Delta k \sim 0.03$~$r.u.$, much less than the DLM case. 
The NM rigid-band expectation corresponds to the vertical black lines, while the spectral peaks for the undoped-NM case are marked by vertical gray lines. For Co-, Ni-, Cu-BFA the peaks lie close to that of rigid-band for electron and hole pockets. Only Zn deviates, see Fig.~\ref{fig-spanv}. This suggests that Zn-BFA has a reduced electron-doping effect and less interaction with Fe and As bands. The reduction of effective $e$/Fe comes from the change in DOS due to the separation of Zn and Fe-host $d$ states well below $E_{F}$, see DOS discussion below.  

\begin{figure}[b]
\includegraphics[width=8cm]{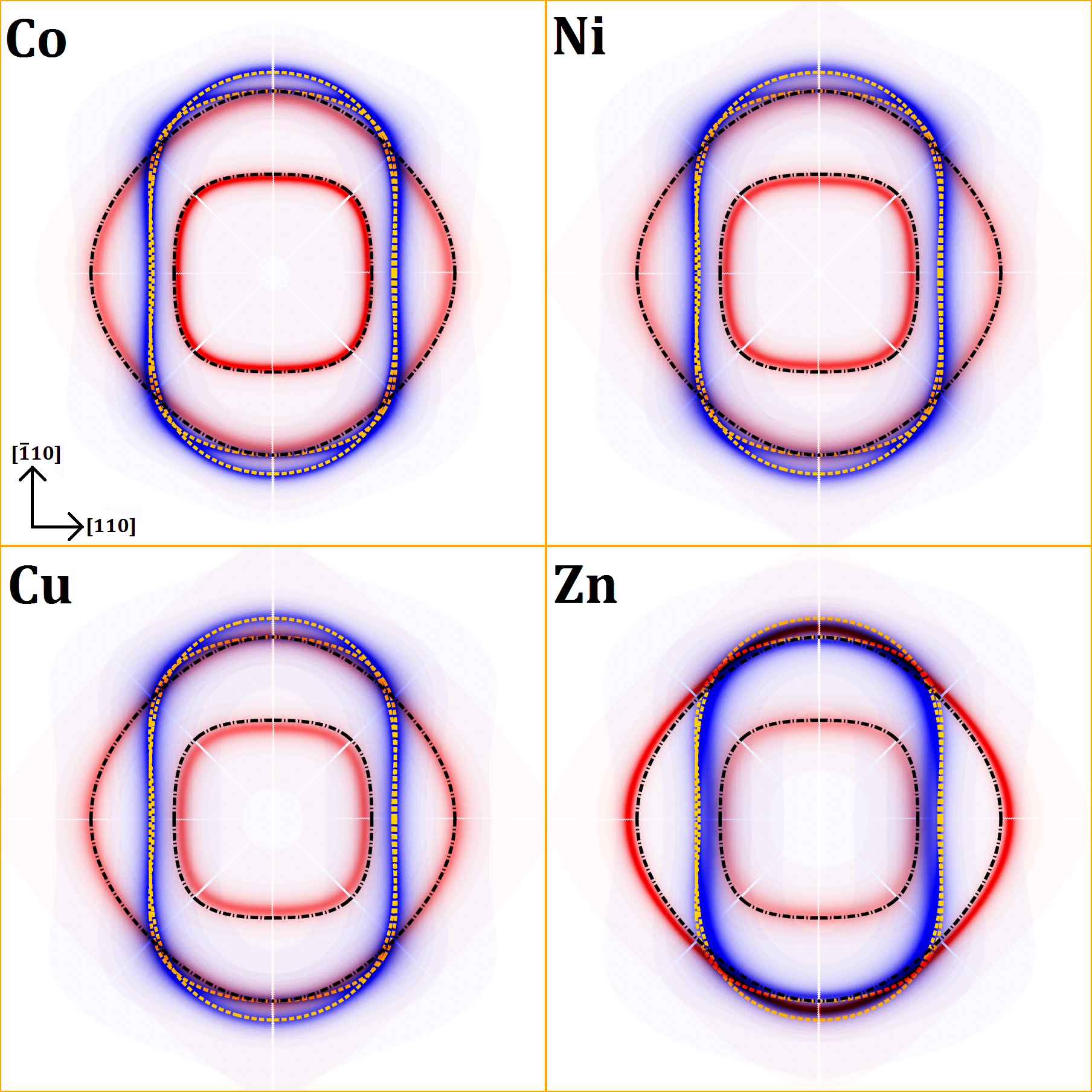}
\caption{(Color online) Overlapped electron (blue) and hole (red) pockets of NM doped compounds at $e$/Fe = 0.10. FS for rigid-band shifted NM at the same $e$/Fe are shown as dashed lines (gold) for electron and dash-dot (black) for hole cylinders.}
\label{fig-dopedfermi}
\end{figure}

These effects are alternatively visible in Fig.~\ref{fig-dopedfermi}, which shows the electron and hole FS at fixed $e$/Fe = 0.10 for Co, Ni, Cu, and Zn doping and compares to that expected from a rigid-band shift from the parent BFA. To show the potential convolution overlap for nesting, the electron surfaces have been shifted to align with hole cylinders for each doped compound. The shift used is the wave-vector connecting X to $\Gamma$, i.e., $\langle\frac{1}{2}\frac{1}{2}0\rangle$. These plots show that the broadening at the Fermi energy is about the same across dopant species for a given $e$/Fe, as expected from Fig.~\ref{fig-spanv}. On doping the holes shrink and electrons grow. This improved ``nesting'' (or overlap) leads to a transverse splitting of the nesting vector along [1$\bar{1}$0], as observed for Co- and Ni-BFA.\cite{PhysRevLett.109.167003} The Zn FS is sharper, indicating longer electron lifetimes. It is visibly shifted from rigid-band expectations, as in Fig.~\ref{fig-spanv}.

Electron states in Fig.~\ref{fig-spanv}(b) [\ref{fig-spanv}(a)] correspond to the vertical $\langle\bar{1}10\rangle$ [horizontal $\langle{1}10\rangle$] direction in  Fig.~\ref{fig-dopedfermi} when transversing from the center. The convolution arises from the entire FS and depends on the broadening and similar widths of spectral features, which increase phase-space overlap volume; but, from the two electron peaks in Fig.~\ref{fig-spanv}(b) and second hole peak in (c) we can make an eye-ball estimate of the incommensurability expected from nesting at E$_F$ from Eq.~(\ref{S2}). Note that in (c) no hole states reflect rigid-band behavior. For Co-doping, the estimate is $0.01$ [$0.03$] in $2\pi/a$ units, spanning that observed value.\cite{PhysRevLett.109.167003} For Ni, it is $0.01$ [$0.03$], again spanning that observed. For Cu, it is near $0$ [$0.02$]. For Zn, it is $-0.02$ [$+0.01$], but the two Zn spectral features are not well separated, smearing the convolution.

Notably, ARPES finds a disagreement between rigid-band lines and the FS of Cu-BFA,\cite{PhysRevLett.110.107007} and no FS changes for Zn-BFA.\cite{PhysRevB.87.201110} However, a DFT study using supercells found a significant shift in the FS of Zn-BFA,\cite{PhysRevLett.108.207003} but the FS shows considerably more broadening than visible here. Thus, there is an apparent discrepancy in electron itinerancy and effective doping between DFT theory and ARPES.
Our calculations too show that Zn FS does not coincide with that of the parent compound, Fig.~\ref{fig-spanv}, and the volume spanned by the electron surface are reduced compared to that expected from rigid-band, see Fig.~\ref{fig-dopedfermi}. In fact, for the $e$/Fe of $0.10$, the effective $e$/Fe is closer to $0.05$ (a 50\% reduction) from direct calculations; an eye-ball estimate from Fig.~\ref{fig-spanv} shows that the Zn spectral peaks are center between the vertical solid lines or the vertical dashed lines Ð which is expected for a rigid-band with $e$/Fe of $0.05$, as calculated. 

A warning to the reader, quantitative agreement with the experimental ARPES spectra from DFT electronic structure  can be more reliably obtained by performing realistic photocurrent calculations that include a proper treatment of the surface electronic structure, energy-dependent matrix elements and lifetime effects, as has been done in KKR.\cite{0305-4608-11-11-027,PhysRevLett.53.2038} For $s$-polarized light, for example, the surface can play only a minor role in photoemission and the measured spectra may follow the DFT quasiparticle dispersion. Otherwise the energy-dependent matrix elements, e.g., from Fermi's ``Golden Rule'' involving photocurrent and the single-site wave-functions, affect the calculated spectra from DFT dispersion. In short, the DFT electronic structure does not necessarily have one-to-one correspondence to that from ARPES, but sometimes it does. So, our above results may all be correct, but, in the future, a more careful comparison is needed with ARPES.


\begin{figure}[]
\centering

\begin{subfigure}
\centering
\includegraphics[width=8cm]{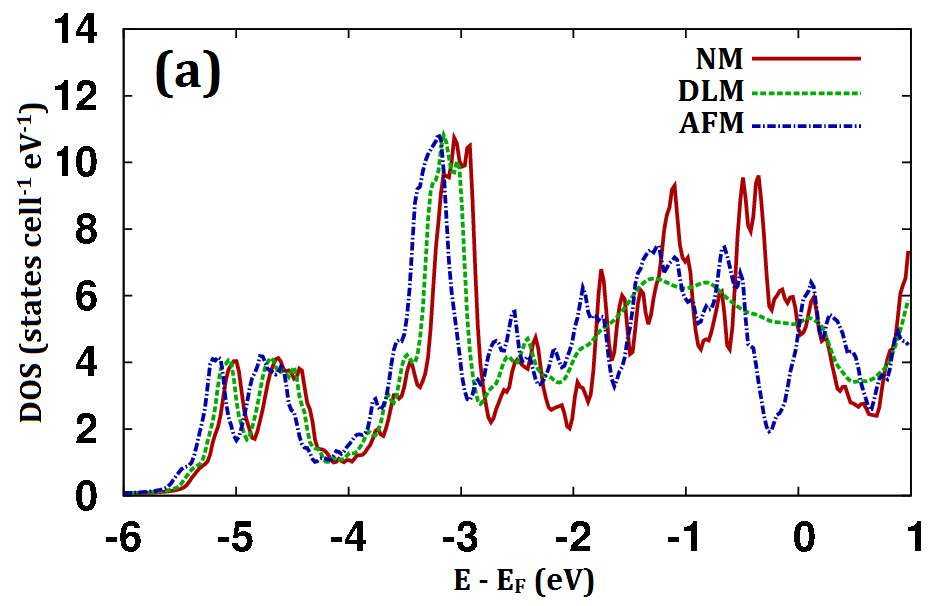}
\end{subfigure}

\begin{subfigure}
\centering
\includegraphics[width=8cm]{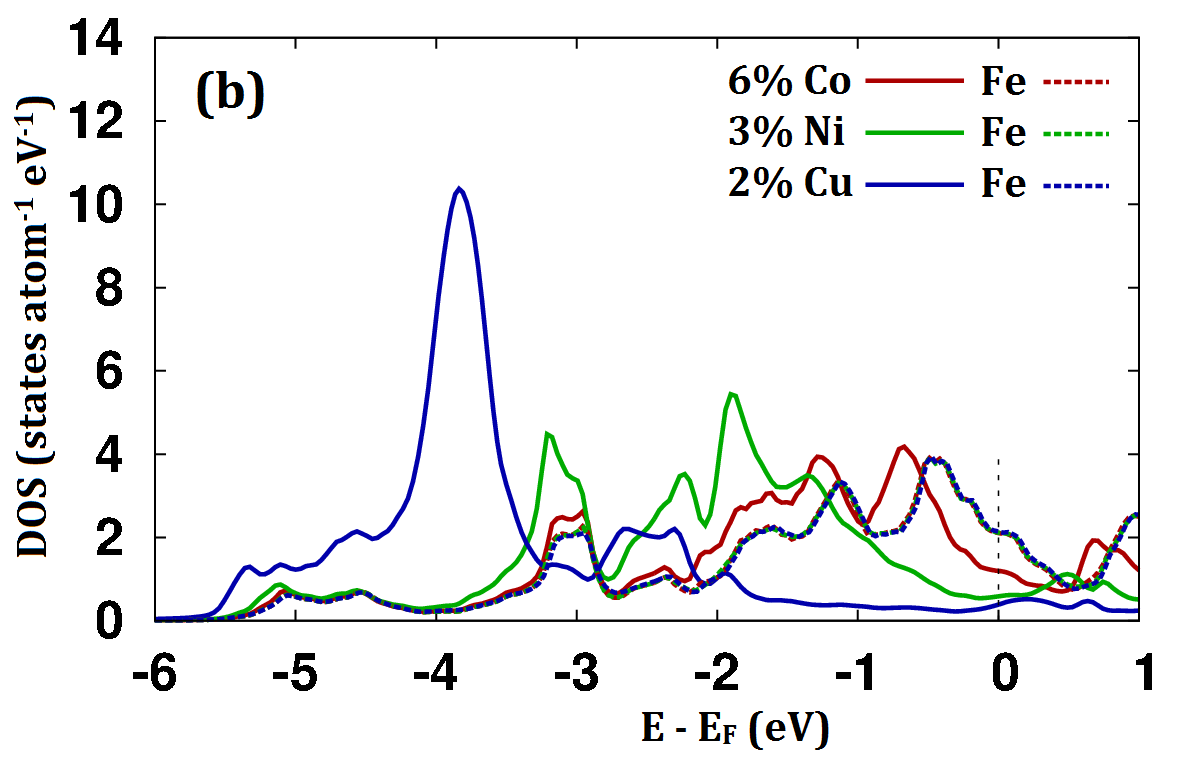}
\caption{(Color online) DOS relative to E$_F$ for (a)  BaFe$_2$As$_2$ NM, DLM, and AFM states, and (b) Ba(Fe$_\text{1-x}$M$_\text{x}$)$_2$As$_2$ M- or Fe- site-projected DOS for fixed $e$/Fe=$0.06$ (i.e., 6\% Co, 3\% Ni, and 2\% Cu). Fe-DOS unaffected by choice of dopant. Zn states are below $-6~e$V with no overlap to Fe-As valence. }
\label{fig-dos}
\end{subfigure}

\end{figure}

\subsection{Density of States and Band Filling}
For BaFe$_2$As$_2$ the valence DOS for NM, DLM, and AFM states are shown in Fig.~\ref{fig-dos}, relative to their respective Fermi energies, E$_F$. The AFM DOS per primitive (i.e., NM) cell are used to ease comparison. From $-6$ to $-3$ $e$V there is strong similarity of the states, but with a shift of E$_F$ due to a pseudo-gap forming below E$_F$ for AFM state and, more weakly, for the DLM state. This shift is +42 m$e$V (DLM) and +126 m$e$V (AFM) relative to the NM. From $-2$ to $1$ $e$V the DLM states are significantly broadened to due local spin disorder. Note the average slope for NM and DLM states near the Fermi level is negative. This can explain the apparent Fermi level shift of the DLM visible in Fig.~\ref{fig-purefermi}. The negative slope and disorder broadening together result in a net reduction in filled states as disorder is turned on. This is compensated by an increased Fermi level. The AFM state shows the opening of a pseudo-gap below the Fermi level, which also explains a large positive shift. The density of states at E$_F$ (i.e., $n(E_F)$) are 5.0, 5.2, and 4.8 states-cell$^{-1}$-$e$V$^{-1}$ for NM, DLM, and AFM, respectively.

For the doped cases of Ba(Fe$_\text{1-x}$M$_\text{x}$)$_2$As$_2$, we focus on the valence DOS for NM states versus M in Fig.~\ref{fig-dos}b. The Fe site-projected DOS do not change for all species M (they clearly lie on top of each other). There is significant overlap of Co and Ni site-projected DOS with Fe-site DOS (common-band behavior), there is clearly a split between states (split-band behavior) on Cu,Zn and Fe (Zn $d$-states are well below $-6~e$V and are not shown). These site projections agree with core-electron spectroscopy. 

The shift for each dopant's $d$-states relative to Fe arises from the increasing $\Delta Z$, where by $\Delta Z_{Cu}=+3$ the $d$-states are no longer in common energy range as Fe. With $\Delta Z_{Zn}=+4$ change in nuclear charge from Fe, the Zn $d$-bands shifts lower in energy creating a split-band (relative to Fe), as will be evident in the DOS, leading to stronger difference in $d$ potentials between Fe and Zn (less so for Cu). 
The common band behavior of Fe and Co,Ni leads to weak impurity scattering and a limited effect on electronic lifetimes and band structure. Conversely, the split band character of Fe and Cu,Zn leads to strong scattering.
These electronic effects are reflected in $\Delta E_{f}$ trends for PM states in Fig.~\ref{fig-energetics}, where both Cu and Zn have positive $\Delta E_{f}$ (unfavorable to mixing with Fe) but Zn less so due to the separation of Zn and Fe-host $d$ states well below $E_{F}$. This changes the overall energetics and  outlines the origin for deviations of Cu,Zn formation energies from Co,Ni in the PM state.


\section{Summary}
In summary, using the all-electron KKR-CPA within DFT, we examined the phase stability, electronic structure, and Fermi-surface evolution of Ba(Fe$_{1-x}$M$_{x}$)$_2$As$_2$ (BFA) with M={Co, Ni, Cu, Zn} for nonmagnetic, paramagnetic, and antiferromagnetic states in high-T tetragonal and low-T orthorhombic structures. Hence, both chemical (alloying) and magnetic (orientational) disorder was addressed. Properties were assessed in terms of additional electrons-per-Fe ($e$/Fe), expected from Hume-Rothery or rigid-band-like behavior. The paramagnetic phase was approximated by a single-site, disordered local moment state that has a finite, randomly oriented moment on each site, which is in contract to the NM state with zero moments. Magnetic effects are pronounced, leading to significant broadening of the Fermi surface and, so, a reduction in coherent carriers; yet, DLM is expected to support the same Fermi-surface nesting effects as from the NM. For the NM state, typically assessed for Fermi-surface nesting instabilities, we find differences versus nominal $e$/Fe in the formation energies, electronic structure, and Fermi-surface properties for Co- and Ni-BFA versus Cu- and Zn-BFA, due to well-known ``split-band'' behavior. Notably, while Cu-BFA deviates from rigid-band in its formation energetics, it continues to follow the rigid-band expectation in the Fermi-surface evolution; but, Zn-BFA does not follow rigid-band in either formation energetics or Fermi surface behavior; we showed that Zn has an effective $e$/Fe that is 50\% of that expected from rigid-band theory due to alloying effects.  This systematic assessment of the electronic properties for all competing states and structures in BFA should help resolve conflicting interpretations based different experiments and theories. Yet, for better comparison to experiment photoemission current calculation using the DFT dispersion would be the best.


\section{Acknowledgements}
Work was supported by the U.S. Department of Energy, Office of Science, Basic Energy Sciences, Materials Science and Engineering Division at Ames (D.D.J.) and through the Center for Defect Physics (S.N.K.), an Energy Frontier Research Center at ORNL. The Ames Laboratory is operated for the U.S. DOE by Iowa State University under contract DE-AC02-07CH11358.


\bibliography{paper.v6}

\begin{thebibliography}{62}
\bibitem [{not()}]{note.entropy}%
  \BibitemOpen
  \href@noop {} {}\bibinfo {howpublished} {Free energy change per atom is the
  formation energy plus the ideal mixing entropy, i.e., $\Delta s = (2/5) k_B T
  [x \log(x) + (1-x)\log(1-x)]$, with annealing T=1000\,$^\circ$C. The 2/5
  arises because Fe/$M$ occupy only 2/5 of all sites.}\BibitemShut {Stop}%
\end{thebibliography}

\end{document}